\documentclass[twocolumns,a4paper]{aa}
\usepackage{epsfig}
\usepackage{natbib} 
\bibpunct{(}{)}{;}{a}{}{,} 
\usepackage{graphicx}
\begin{document}
\title{High speed magnetized flows in the quiet Sun}
\author{C. Quintero Noda\inst{1,2} \and J.M. Borrero\inst{3} \and D. Orozco Su\'arez\inst{1,2} \and B. Ruiz Cobo\inst{1,2}}
\institute{Instituto de Astrof\'isica de Canarias, E-38200, La Laguna, Tenerife, Spain.\ \email{cqn@iac.es} \and
Departamento de Astrof\'isica, Univ. de La Laguna, La Laguna, Tenerife, E-38205, Spain \and
Kiepenheuer-Institut f\"ur Sonnenphysik, Sch\"oneckstr. 6, D-79104, Freiburg, Germany}

\date{Received May 2014/Accepted July 2014}

\abstract 
{We analyzed spectropolarimetric data recorded with Hinode/SP in quiet-Sun regions located at the disk center. We found single-lobed Stokes $V$ profiles showing highly blue- and red-shifted signals. Oftentimes both types of events appear to be related to each other.}
{We aim to set constraints on the nature and physical causes of these highly Doppler-shifted signals, as well as to study their spatial distribution, spectropolarimetric properties, size, and rate of occurrence. Also, we plan to retrieve the variation of the physical parameters with optical depth through the photosphere.}
{We have examined the spatial and polarimetric properties of these events using a variety of data from the Hinode spacecraft. We have also inferred the atmospheric stratification of the physical parameters by means of the inversion of the observed Stokes profiles employing the Stokes Inversion based on Response functions (SIR) code. Finally, we analyzed their evolution using a time series from the same instrument.}
{Blue-shifted events tend to appear over bright regions at the edge of granules, while red-shifted events are seen predominantly over dark regions on intergranular lanes. Large linear polarization signals can be seen in the region that connects them. The magnetic structure inferred from the time series revealed that the structure corresponds to a $\Omega$-loop, with one footpoint always over the edge of a granule and the other inside an intergranular lane. The physical parameters obtained from the inversions of the observed Stokes profiles in both events show an increase with respect to the Harvard-Smithonian reference atmosphere in the temperature at $\log\tau_{500} \in (-1, -3)$ and a strong magnetic field, $B \ge 1$ kG, at the bottom of the atmosphere that quickly decreases upward until vanishing at $\log\tau_{500} \approx -2$. In the blue-shifted events, the line of sight velocities change from upflows at the bottom to downflows at the top of the atmosphere. Red-shifted events display the opposite velocity stratification. The change of sign in line of sight velocity happens at the same optical depth in which the magnetic field becomes zero.}
{The physical mechanism that best explains the inferred magnetic field configuration and flow motions is a siphon flow along an arched magnetic flux tube. Further investigation is required however, as the expected features of a siphon flow cannot be unequivocally identified.}
\keywords{Sun: magnetic fields – Sun: photosphere – Sun: granulation}
\titlerunning{High speed magnetized flows in the quiet-Sun}
\authorrunning{Quintero Noda et al.}
\maketitle

\section{Introduction}

Stokes profiles are used to characterize the properties of partially polarized light: the Stokes $I$ profile gives information about the intensity (total number of photons), $Q$ and $U$ about the linear polarization, and $V$ about the circular polarization  \citep{Rees1989,Landi1992,delToro2003}. In an atmosphere at rest, harboring a vertically homogeneous magnetic field, under the assumption of local thermodynamic equilibrium (LTE), and within the Zeeman regime, Stokes $V$ is expected to be strictly antisymmetric with respect to the zero-crossing wavelength, and remains unshifted with respect to the position of the Stokes $I$ line core. Under these assumptions the so-called area asymmetry (or $\delta A$), that is, the wavelength integral of the signed Stokes $V$ signal normalized to its total area, is zero. A non-vanishing $\delta A$ is a necessary and sufficient condition for the existence of vertical variations in the line of sight component of the velocity and/or magnetic field vector \citep{Landolfi1996}. Recent observations at high spatial resolution have revealed Stokes $V$ profiles with distinct asymmetries in active regions \citep{Balasu1997}, plages \citep{MartinezPillet1997,Sigwarth1999} and the quiet Sun \citep{Vitichie2011,Sainz2012}. In the case of extreme asymmetries, one of the lobes of Stokes $V$ is missing, and therefore the $\delta A=1$. These are commonly referred to as single-lobed Stokes $V$ profiles.

Several theoretical models featuring large gradients in velocity and magnetic field have been suggested to explain the existence of single-lobed Stokes $V$ profiles. \cite{Steiner2000} proposed that these profiles can be produced when the line of sight (LOS) passes through a magnetopause, i.e., a separatrix layer dividing the atmosphere in two regions with different magnetic fields. Magnetopauses occur frequently in the solar atmosphere, with examples being the canopies created by expanding flux tubes in the network \citep{Grossmann2000} and the interfaces between different magnetic components in sunspot penumbrae \citep{Schlichenmaier1998}.

In the present work, we focus on the analysis of highly asymmetric ($\delta A \rightarrow 1$) and highly Doppler-shifted Stokes $V$ signals in the quiet Sun detected with the spectropolarimeter SP \citep{Lites2013_SP} on board the Hinode spacecraft \citep{Kosugi2007,Tsuneta2008}. Although these profiles are probably included in the extensive analysis carried out by \cite{Sainz2012}, these authors do not focus on those profiles, which besides having $\delta A \rightarrow 1$, are also strongly Doppler-shifted. The origin of Stokes $V$ profiles featuring both properties still remains under debate.

\cite{BellotRubio2001}, \cite{Shimizu2008}, \cite{Nagata2008} and \cite{Fischer2009} concluded that the cause of the strong red-shifted signals they found is convective collapse inside a magnetic flux tube. This process, which was first proposed on theoretical grounds by \cite{Parker1978}, \cite{Webb1978}, and \cite{Spruit1979}, makes the plasma inside a magnetic flux tube unstable, cool down, and fall along the field lines, sometimes at supersonic speeds. Meanwhile, the flux tube narrows and the magnetic field inside intensifies. This happens until a critical magnetic field (about $1-2$ kG), capable of suppressing the instability, is reached. The downflowing plasma can hit denser layers beneath it and rebound upward. This upflow could explain the extremely blue-shifted signals observed by \cite{SocasNavarro2005}.

Another mechanism that could produce strong up- and downflows is magnetic reconnection \citep{Parker1963}. The change in the topology of the magnetic field lines, favored by the low electrical conductivity of the photospheric layers, can produce two opposite jets that could reach supersonic velocities. Sweet-Parker reconnection, as modeled by \cite{Chae2002} and observed by \cite{Litvinenko2007}, could produce jets with outflowing speeds in the range of 3-10 km/s. This mechanism was proposed by \cite{QuinteroNoda2014} to explain the detected strong downflows embedded in a heated environment where the magnetic field intensity decreases during the process. \cite{Borrero2013} also invoked this scenario to explain the physical parameters inferred through the inversion of the strong Doppler-shifted profiles reported in \cite{Borrero2010}.

Last but not least, the siphon flow mechanism is also capable of producing highly Doppler-shifted signals. This phenomenon arises in a flux tube that connects magnetic elements of opposite polarity provided that there is a difference in the magnetic field between both footpoints. The difference in the magnetic field creates a gas pressure imbalance that accelerates the plasma from the high-pressure point toward the low-pressure footpoint. Under this mechanism, the velocities can also reach supersonic values \citep[][and references therein]{Montesinos1993,Rueedi1992}.

In the following sections we will study the polarimetric properties of two Fe {\sc i} lines observed by Hinode/SP in those regions where the Stokes $V$ profiles are both highly blue-shifted and present only one lobe (i.e., area asymmetry $\delta A \rightarrow 1$). Then, we will follow their vertical trace through the analysis of the chromospheric Ca {\sc ii h} spectral band. Finally, we will infer the atmospheric stratification of the physical parameters using the Stokes Inversion based on Response functions (SIR) \citep{RuizCobo1992} code. We aim to reveal the nature of these strongly asymmetric and highly Doppler-shifted signals as well as to reveal their underlying physical mechanism.

\section{Observations and data analysis}
\label{observations}

The polarimetric data used in this paper were acquired with the spectropolarimeter \citep[SP ; ][]{Lites2013_SP} on board the Hinode spacecraft \citep{Kosugi2007}. We selected a data set with a field of view of $328^{\prime\prime}\times154^{\prime\prime}$ recorded at disk center on March 10$^{\rm th}$, 2007. The SP instrument measures the Stokes vector of the Fe~{\sc i} line-pair at 630 nm with a spectral and spatial sampling of 2.15 pm pixel$^{-1}$ and 0.16$^{\prime\prime}$ ({\it normal map}), respectively. The exposure time is 4.8 seconds per slit position, making it possible to achieve a noise level of $1.1\times10^{-3}~I_{c}$ in Stokes $V$ and $1.2\times10^{-3}~I_{c}$ in Stokes $Q$ and $U$. Here $I_{c}$ refers to the mean continuum intensity in the granulation.

The second data set is a time series observation, based on a raster scan of 18 slits with 1.6 seconds of exposure time per slit position. It spans three hours of observation on September 25$^{\rm th}$, 2007 ($\mu$=1). In this case, the field of view is $2.9^{\prime\prime}\times38^{\prime\prime}$ and the final time cadence is 36 seconds. This data set belongs to the Hinode Operation Plan 14, entitled Hinode/Canary Islands campaign, and it allows us to analyze the time evolution of small magnetic structures. Although the noise level is slightly higher ($\approx 2\times 10^{-3}~I_{c}$) than in the first data set, it is still good enough to study polarization signals above the $10^{-2}~I_{c}$ level.

We also have at our disposal co-aligned CN and Ca~{\sc ii~h} broadband images recorded by the Broadband Filter Imager (BFI) instrument on Hinode \citep{Tsuneta2008}. The BFI images were acquired simultaneously with the second dataset (time series). The CN filter is located at 383.3 nanometers (filter width of 0.52 nm FWHM) and conveys information about the photosphere. The Ca~{\sc ii~h} filter is located at 396.85 nm (filter width of 0.22 nm FWHM) and contains information from both the photosphere and the chromosphere \citep[see][]{pietarila2009}. The exposure times were 0.1 s and 0.3 s, respectively.

Data from CN and Ca~{\sc ii~h} spectral bands consist of a series of images of the same field of view, taken every 30 seconds. Because of the difference between the cadence of the acquisition of these images and the cadence of each Hinode/SP raster (36 seconds), we choose the broadband images that are closest in time to the ninth slit position of the raster scan performed by the SP instrument. With this, we guarantee that the time delay between any SP raster and the BFI images is never larger than 18 seconds.

\begin{figure*}
\centering
\includegraphics[width=18cm]{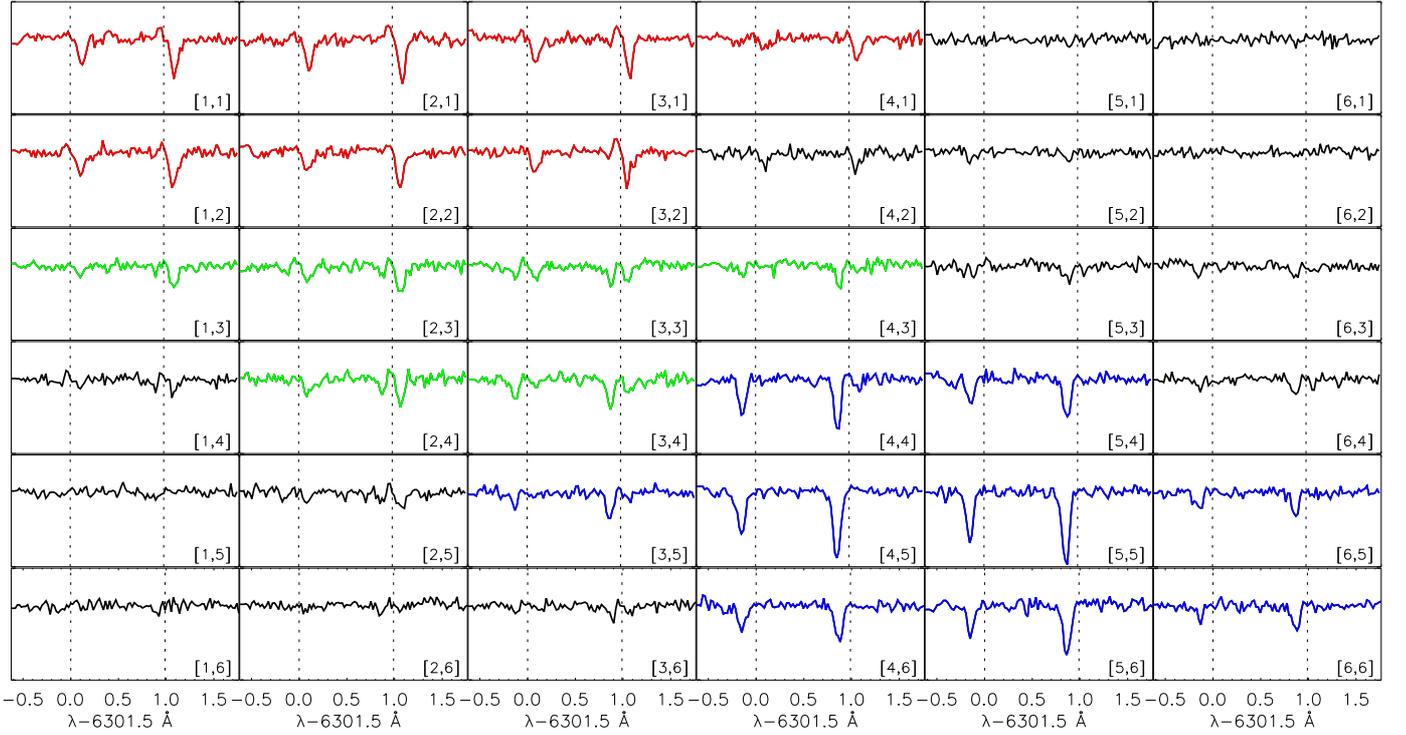}
\vspace{+20pt}
\caption{\footnotesize Typical spatial distribution of the circular polarization profiles (Stokes $V$) around the events detected on the {\it normal} map. Each panel displays adjacent profiles from the two iron lines located at 6301.5 and 6302.5 \AA \ measured by Hinode/SP. The vertical scale ranges from -4\% to +2\% of $I_c$. Blue  denotes the detected single-lobed blue-shifted profiles. Green shows pixels with significant linear polarization signals ($\ge 0.01~I_{c}$). Finally, the red lines indicate pixels with single-lobed red-shifted Stokes $V$ profiles. Vertical lines indicate the rest center wavelength of each line.} 

\label{spa1t}
\end{figure*}

\begin{figure*}
\centering
\includegraphics[width=18cm]{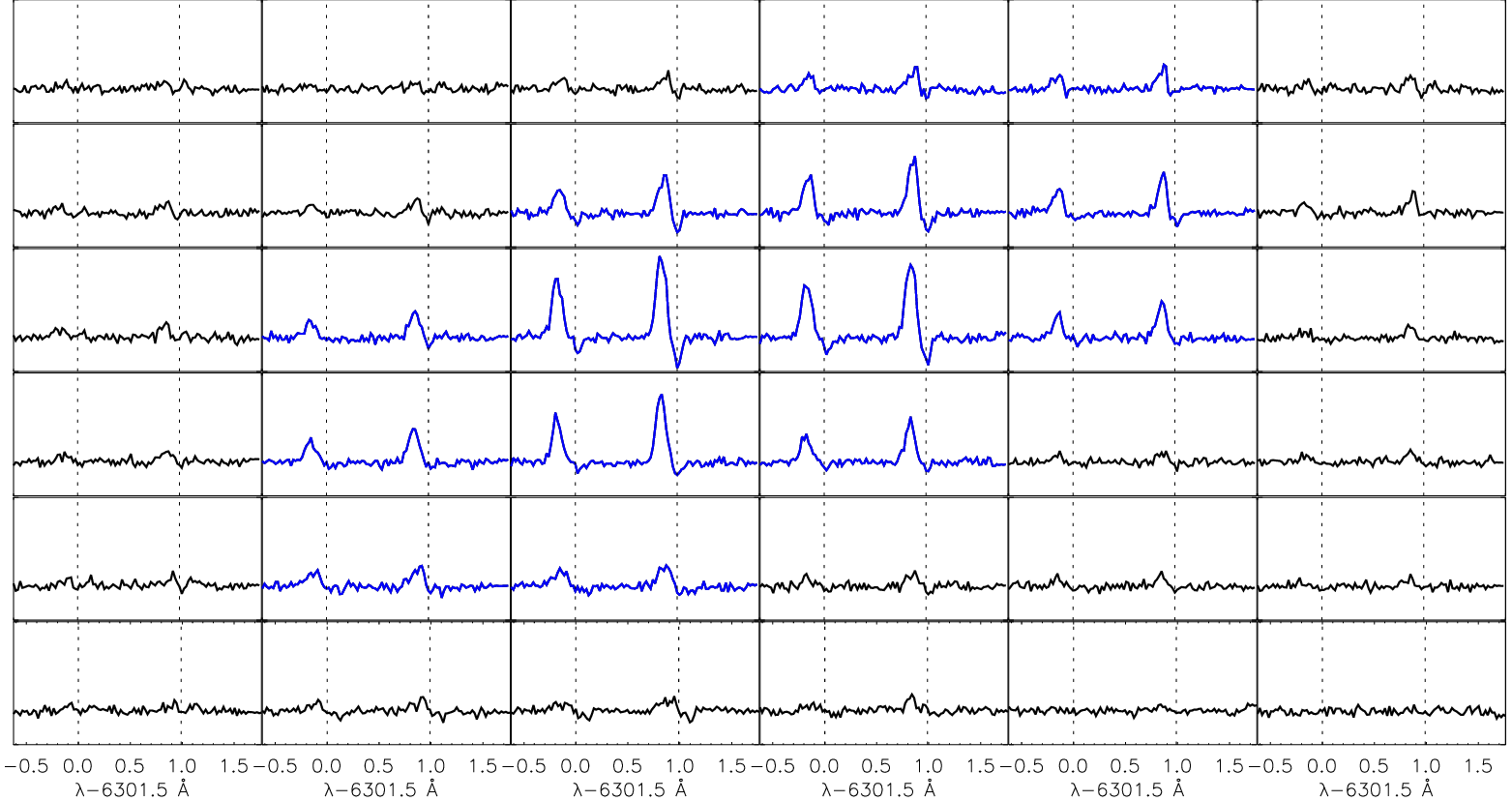}
\vspace{+20pt}
\caption{Same as Figure \ref{spa1t}, except showing an example of a blue-shifted case that appears unrelated to any red-shifted signal. The vertical scale goes from -3\% to +8\% of $I_c$.} 
\label{spa2}
\end{figure*}

The events of interest in this work are those presenting strong Doppler-shifted Stokes $V$ profiles with a single lobe. We have followed the method described in \cite{MartinezPillet2011a} to find them, that is, examining the Stokes $V$ signal at $\pm 272$~m\AA \ from the center of the Fe {\sc i} 6302.5 \AA \ line (i.e., on the continuum) and generating Stokes $V$ \textit{continuum} magnetograms from the red and blue sides of that line. Then, we analyzed these magnetograms looking for pixels with Stokes $V$ signals above 0.5\% of the mean continuum intensity of the map ($\ge 0.05~I_{c}$). As the detected events were grouped in patches of different sizes, we decided to define a minimum area of detection of two pixels to prevent isolated pixels from entering the analysis. Isolated spikes can be produced by cosmic rays or data compression issues. 

\begin{figure*}
\centering
\includegraphics[width=29.5cm]{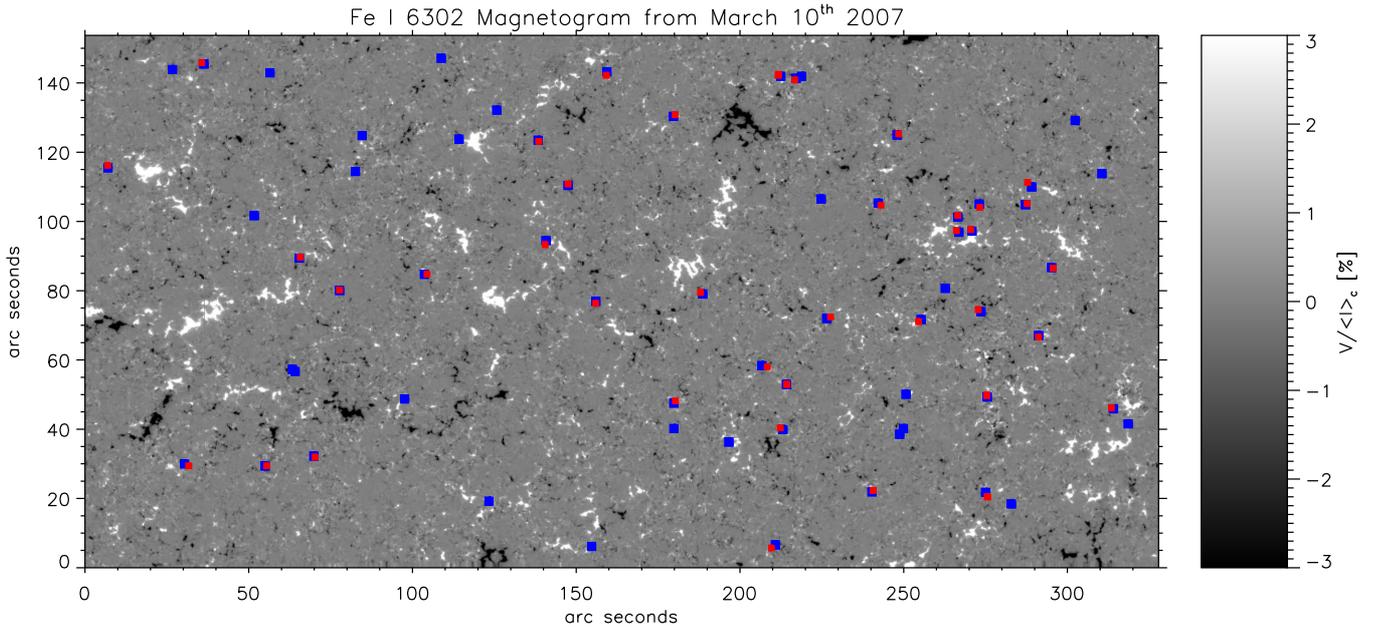}
\caption{Spatial distribution of the blue and red cases over the Fe {\sc i} 6302.5~\AA \ magnetogram from the Hinode/SP \textit{normal} map. Blue squares indicate the location of the blue events, while red squares designate the position of the red events.} 
\label{magn}
\end{figure*}

\section{General properties of the single-lobed Stokes $V$ profiles}

\subsection{Polarimetric description}\label{profiles_sec}

With the selection criteria described in Section~\ref{observations}, we have found 398 strongly blue-shifted Stokes $V$ profiles belonging to 64 different events and 158 highly red-shifted Stokes $V$ profiles corresponding to 46 events. Hereafter, we will refer to these events as {\it blue} and {\it red} cases, respectively. Although not demanded by our selection criteria, all the selected circular polarization profiles happen to be highly asymmetric, $\delta A \rightarrow 1$, featuring one single lobe. In addition, the intensity profiles (Stokes $I$) also possess some common features. For instance, Stokes $I$ in the blue cases, present a mean continuum signal above one, indicating that they are located over bright regions on the continuum map (granular edges). The red cases, however, show the opposite behavior and display a mean continuum intensity below one, thus appearing preferably over dark regions on the continuum map (intergranular lanes). We have also found that the outer-blue wing of Stokes $I$ is strongly pushed toward shorter wavelengths in the blue cases. Likewise, the outer red wing of Stokes $I$, in the red cases, presents a strong shift toward larger wavelengths. This produces strongly asymmetric intensity profiles that are indicative of large depth variations in the LOS component of the velocity.

In addition, we have searched for blue- or red-shifted circular polarization signals inside a $2\times2\arcsec$ region centered in every detected event, and we found that 73\% of the blue cases are related to a red event. The rest of the blue cases, 27\% of the total, seem to be isolated events. Interestingly, red cases never appear isolated. The distance between blue and the associated red cases varies but, whenever both events are close enough, they are connected by pixels that present strong linear polarization.  In those in-between regions with large levels of linear polarization, Stokes $V$ is also highly asymmetric ($\delta A \rightarrow 1$), but in this case the profiles are not single lobed. Instead, they present two lobes of the same sign. Figure~\ref{spa1t} shows a typical example of the spatial distribution of the Stokes $V$ profiles whenever a blue event appears related to a red event. This figure is composed of $6\times6$ panels representing adjacent pixels. Each panel displays the Stokes $V$ profiles of both iron lines,  Fe {\sc i} 6301.5 and 6302.5~\AA. Dotted vertical lines represent the rest center wavelength. In every panel, the vertical scale goes from -4\% to +2\% of I$_c$. We plotted in blue the Stokes $V$ profiles that we classify as belonging to a blue event, while red indicates the profiles with Stokes $V$ belonging to a nearby red event. Green profiles in between the blue and red profiles correspond to the Stokes $V$ profiles where the linear polarization is above 1\% of I$_c$. We note that the closest pixels to the green profiles that display Stokes $V$ blue- or red-shifted signals, also show a non-negligible amount of linear polarization, albeit below the 1\% level. The rest of the pixels in blue or red (e.g., panels on the top leftmost, $[1,1]$, and bottom rightmost, $[6,6]$, part in Fig.~\ref{spa1t}) do not possess a significant amount of linear polarization. We emphasize that the polarization signals in these observations, with circular and linear polarization levels higher than $0.03~I_{c}$ and $0.01~I_{c}$, respectively, are much stronger than those in the quiet-Sun internetwork \citep[cf. Fig.~4 in][]{borrero2013clv}.

An interesting property of the events in this configuration (Fig.~\ref{spa1t}) is that the sign of the single lobe in Stokes $V$ is the same for the red and blue cases: if one of the patches displays negative single-lobed profiles, then the associated event will also show negative Stokes $V$ profiles but with the opposite wavelength shift, and vice-versa (e.g., compare panels $[2,2]$ and $[5,5]$). We also see that the missing lobe is always the one closer to the rest wavelength (vertical dotted lines in each panel). These two facts seem to indicate that the circular polarization profiles in the blue and red cases are produced by magnetic fields of opposite polarity that harbor plasma flows along opposite directions. This can be realized by picturing circular polarization profiles where the missing lobe, in both red and blue cases, would be present close to the rest wavelength (vertical dotted lines). In addition, we observe that the Stokes $V$ profiles in the region that connects the red and blue events (green profiles in Fig.~\ref{spa1t}) are also highly asymmetric. Unlike the blue and red Stokes $V$ profiles, these profiles are not single-lobed, but rather they possess two lobes of the same sign (i.e., panel $[3,3]$).

The properties described above point to a configuration of the magnetic field in the form of a magnetic loop, where the footpoints possess a magnetic field of opposite polarities and velocities of opposite signs. The detection of linear polarization signals between footpoints further supports this picture. Interestingly, this scenario could also explain the observed asymmetries in Stokes $V$. The circular polarization profiles at the footpoints would have single-lobed shapes because the observer sees two different atmospheres along the line of sight: one with magnetic field and the other without it \citep{Viticchie2012}. On the other hand, the Stokes $V$ profiles in the region with large linear polarization signals (green colors in Fig.~\ref{spa1t}) can be ascribed to the limited spatial resolution of the telescope, either produced by the contamination of the point spread function the telescope introduces in each pixel \citep{vanNoort2012} or by the presence of two magnetic fields with opposite polarities on the same resolution element \citep{Sigwarth2001}. For instance, Stokes $V$ in panel $[3,3]$ can be thought as being produced by a mixture of the Stokes $V$ profiles from panels $[2,2]$ and $[4,4]$. In spite of all the information gathered by the visual inspection of the profiles, it is so far not possible to establish whether the magnetic loop corresponds to a $\Omega$-loop or a U-loop. We will answer this question in the following sections.

Finally, Figure~\ref{spa2} illustrates an example of the remaining 27\% of detected blue-shifted events, where no red-shifted counterpart could be identified within a $2\times2\arcsec$ region around it. It was not possible to detect any significant linear polarization signals there either. In this figure, the vertical scale ranges from -3\% to +8\% of I$_c$. The blue-shifted single-lobed Stokes $V$ profiles in Figure~\ref{spa2} are very similar to those displayed in Figure~\ref{spa1t}. The only discernible difference is that the amplitude of the blue-shifted Stokes $V$ lobe in the isolated events is generally larger than the amplitude of the blue-shifted Stokes $V$ lobe when the blue event is accompanied by a red event.

\subsection{Spatial distribution and rate of occurrence}
\label{spatial_dist}

To analyze the spatial distribution of the detected events we have plotted in Figure \ref{magn} the locations of all blue and red cases on a magnetogram constructed by integrating the signed circular polarization signals across the Fe {\sc i} 6302.5 {\AA} spectral line. Blue squares mark the position of the blue cases, while red squares correspond to the red cases. In order to make them visible, the size of each square has been magnified with respect to the real area occupied by each event. The locations of these events seem to be related to regions where the magnetic field is concentrated. They are close to network areas and rarely appear on the internetwork. 

These events appear in patches of different sizes over the entire map. The mean areas are 6.2$\pm$3.1 and 3.5$\pm$2.1 pixels, for blue and red cases respectively. The smaller area of the latter could explain why we detect fewer of them, 46 red events compared to 64 detected blue events. To obtain the rate of occurrence of these events we divide the number of detected cases, 64 blue cases and 46 red cases, by the total area of the map ($328^{\prime\prime}\times154^{\prime\prime}$). This provides a rate of occurrence of $1.2\times10^{-3}$ blue cases per arcsec$^{2}$ and $9.1\times10^{-4}$ red cases per arcsec$^{2}$. These values indicate that we would find nearly 12 different blue events, or nine red cases, in a FOV of $100\times100$ arcsec$^{2}$.

\subsection{Analysis of the time series}
\label{time_evol}

\begin{figure*}
\centering
\includegraphics[width=15.cm]{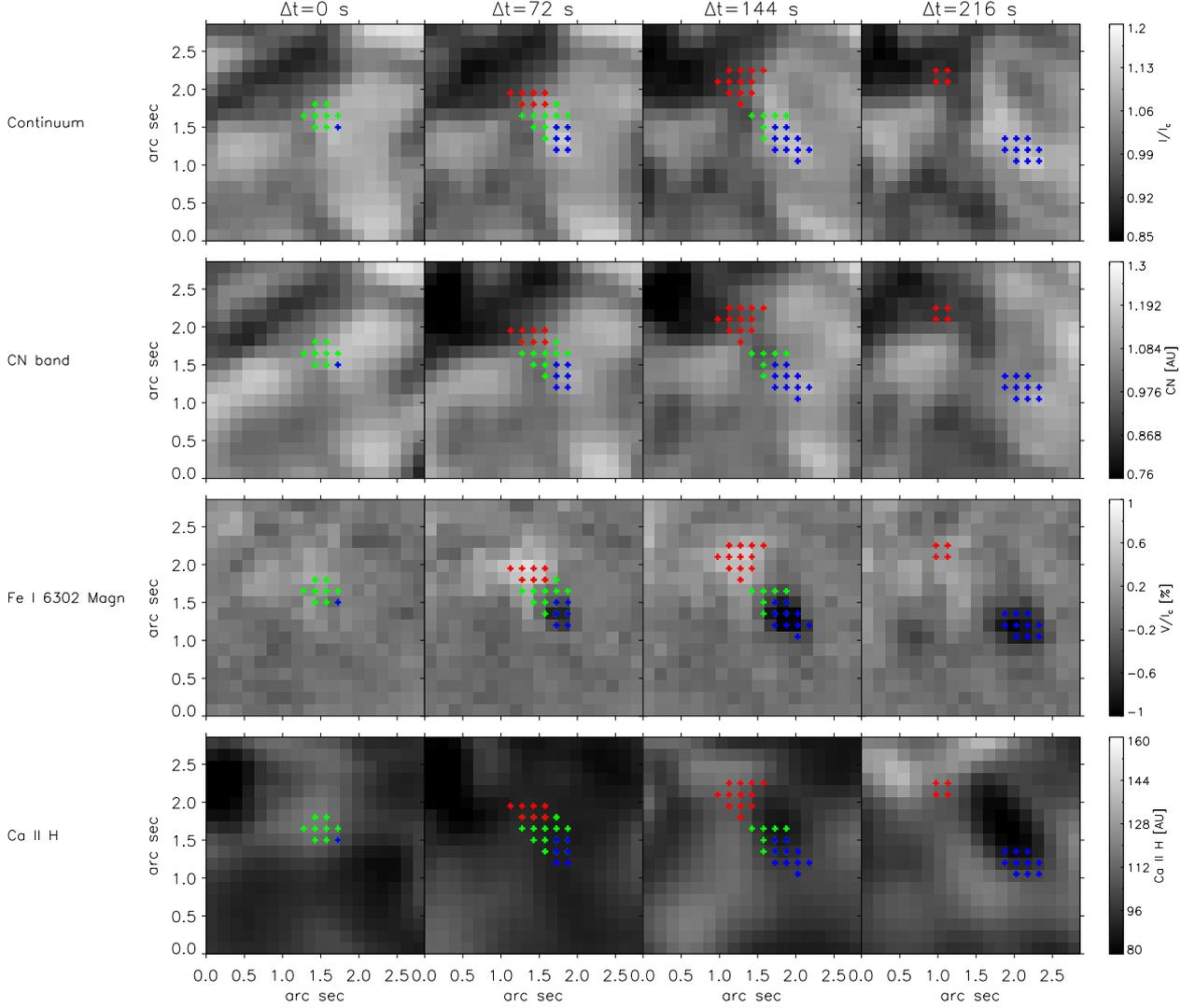}
\caption{Evolution of a single event detected with the Hinode/SP time series. Each row corresponds to a different time in its evolution. The first column displays the continuum signal from Hinode/SP, the second column shows the CN broadband image from Hinode/BFI, the third column displays the Fe {\sc i} 6302.5 \AA \ magnetogram from Hinode/SP, and the fourth column shows the Ca {\sc ii h} broadband images from Hinode/BFI. We have also highlighted the pixels with blue- and red-shifted single-lobed Stokes $V$ profiles with blue and red crosses, respectively. Pixels with linear polarization signals above $0.01~I_{c}$ are indicated with green crosses.} 
\label{chromos}
\end{figure*}

We complement the information presented in Sections~\ref{profiles_sec}, and ~\ref{spatial_dist} using the time series from Hinode/SP described in Section~\ref{observations}. In the three hours of observations available we found five cases that are analogous to those found in the {\it normal} map (see Fig.~\ref{magn})\footnote{Considering the total area spanned by the time series and the rate of occurrence determined in Sect.~\ref{spatial_dist} a total of about 40 events should have been detected. The fact that we see ten times fewer might indicate that the slit of the raster scan probably sit on an internetwork region, where many fewer events are usually seen (see Fig.~\ref{magn})}. Because of the small FOV  ($2.9^{\prime\prime}\times38^{\prime\prime}$) it is difficult to detect any blue case, along with its associated red event, that falls into the small scanned region during the entire lifetime of the event. Out of five detected events, only one of them remains within the field of view throughout its evolution. In Figure~\ref{chromos}, we analyze the time evolution of this particular event using Hinode/SP scans as well as CN and Ca {\sc ii~h} broadband images from Hinode/BFI. The event lasts nearly ten minutes. From top to bottom each row displays the continuum intensity from Hinode/SP, CN broadband images, Fe {\sc i} 6302.5 \AA \ magnetogram (obtained in the same way as Fig.~\ref{magn}), and Ca {\sc ii h} broadband images. Each of the four columns correspond to a different time step in the event's evolution: $\Delta t=0$, 72, 144 and 216 seconds. Following the same convention used in Fig.~\ref{spa1t}, we have also marked in blue the pixels harboring blue-shifted single-lobed Stokes $V$, in red the pixels displaying red-shifted single-lobed Stokes $V$, and in green the pixels with a linear polarization above 1\% of $I_{c}$.

The left most column in Fig.~\ref{chromos} corresponds to the initial detection of the event. At this time, the event displays only a patch of large linear polarization signals and almost no blue- or red-shifted Stokes $V$ profiles. The pixels where the linear polarization is above 1\% of $I_{c}$ exhibit $Q$-like circular polarization profiles characterized by two lobes of the same sign (similar to those in Fig.~\ref{magn}; panel $[3,3]$). After 72 seconds (second column) two patches of blue- and red-shifted single-lobed Stokes $V$ profiles developed and they appear connected through the region of enhanced linear polarization. The third and fourth columns show that the patches of blue- and red-shifted single-lobed circular polarization profiles separate, while the pixels with high linear polarization signals vanish. The footpoints keep moving away from each other for the remaining of the event's lifetime (10 min).  According to the Fe {\sc i} continuum and CN broadband images (first and second rows), the blue- and red-shifted patches lie over bright and dark regions, respectively. The Fe {\sc i} 6302.5 \AA \ magnetogram (third column in Fig.~\ref{chromos}) further demonstrates (see Sect.~\ref{profiles_sec}) that these two patches appear over magnetic fields of opposite polarities. Finally, we do not find any traces of these events higher up in the atmosphere (Ca {\sc ii h} images on the bottom row in Fig.~\ref{chromos}) at any time in their evolution.

As already mentioned in Sect.~\ref{profiles_sec} (where we studied instantaneous images of these events) the observed configuration, as seen from its temporal evolution, is also compatible with a magnetic loop where the linear polarization connects two footpoints with opposite velocities and magnetic fields of opposite polarity. Since the linear polarization precedes the appearance of the footpoints of the loops, the time evolution could correspond to either an emerging $\Omega$-loop or to a submerging U-loop. These kinds of events have been previously reported in \cite{MartinezGonzalez2007,Centeno2007,MartinezGonzalez2009,Ishikawa2010,Viticchie2012}. All of them describe the structure as two footpoints of opposite Stokes $V$ signals connected by linear polarization, but only \citep{Ishikawa2010} found opposite velocities, and only \cite{Viticchie2012} detected single-lobed Stokes $V$ profiles.

\begin{figure*}
\centering
\includegraphics[width=18.4cm]{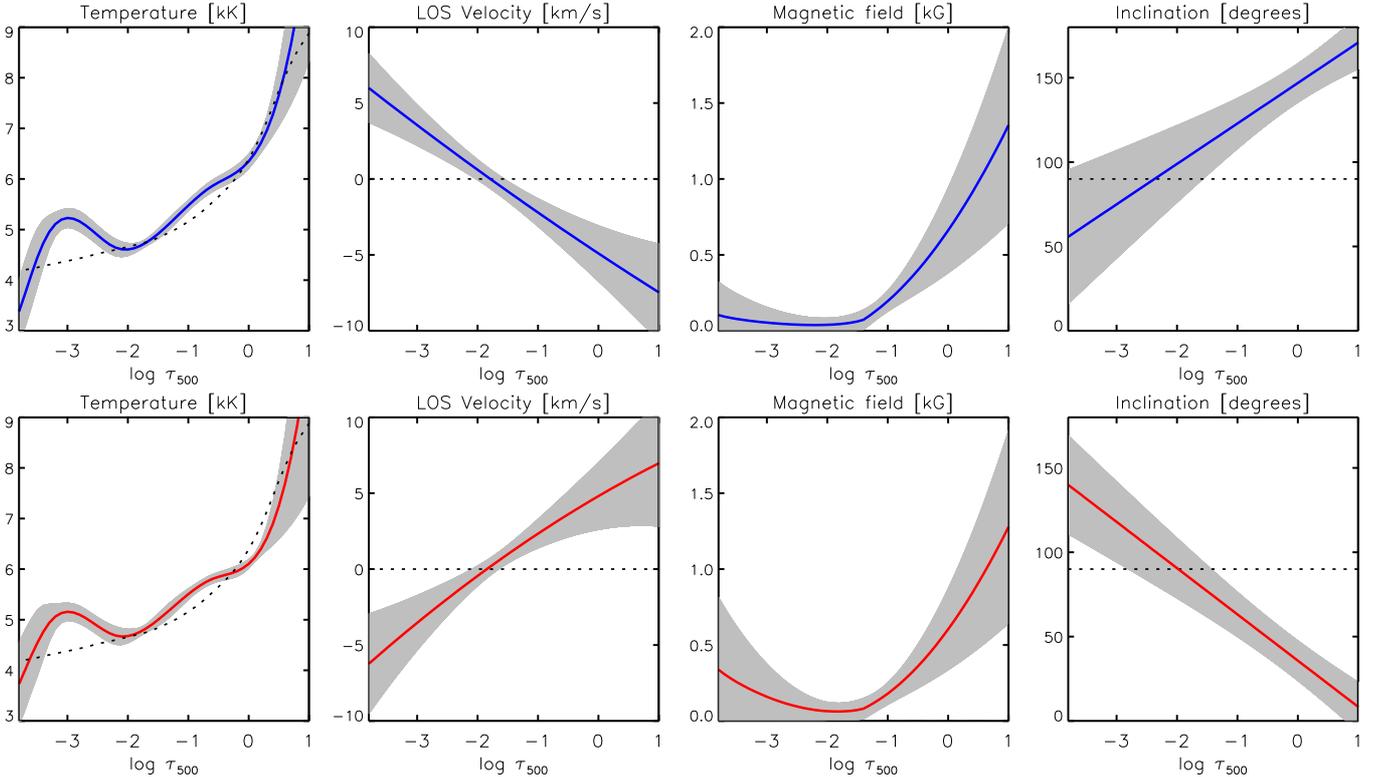}
\caption{Mean atmospheric stratifications obtained from the inversion of the selected pixels in the Hinode/SP \textit{normal} map (Fig.~\ref{magn}). The first and second rows represent the results for blue and red events, respectively. Color coding follows this designation. The first column displays the temperature $T$, second column the LOS velocity ${\rm v}_{\rm LOS}$ with positive values representing downflows, third column the magnetic field intensity $B$, and fourth column the inclination of the magnetic field $\gamma$. Gray areas indicate the deviation from the mean stratifications obtained from the 358 pixels belonging to blue events and 158 pixels belonging to red events (see Sect.~\ref{profiles_sec}).}
\label{mean}
\end{figure*}

\section{Stokes profiles inversions}

\subsection{Configuration}\label{conf}

The SIR code \citep[Stokes inversion based on Response Functions;][]{RuizCobo1992} allows us to infer the optical-depth dependence of the atmospheric parameters at each pixel through the inversion of the observed Stokes profiles. In our inversions we employ a single magnetic component parametrized by: seven nodes in the temperature $T(\tau_{500})$\footnote{The parameter $\tau_{500}$ refers to the optical depth evaluated at a wavelength where there are no spectral lines (continuum). In our case this wavelength is 500 nm.}, three in the line of sight (LOS) component of the velocity ${\rm v}_{\rm LOS}(\tau_{500})$, three in the magnetic intensity $B(\tau_{500})$, two for the inclination of the magnetic field with respect to the observer's LOS $\gamma(\tau_{500})$, and one for the azimuthal angle of the magnetic field in the plane perpendicular to the observer's LOS $\phi(\tau_{500})$. Variables such as micro- and macro-turbulence are fixed to zero and not inverted. At each iteration step the synthetic profiles are convolved with the spectral transmission profile of Hinode/SP \citep{Lites2013_SP}. Since each node corresponds to a free parameter during the inversion, our model includes a total of 16 free parameters. The total number of data points is 448 distributed across 112 wavelength positions for each of the four Stokes profiles.

The nodes in $B(\tau_{500})$, $\gamma(\tau_{500})$, and ${\rm v}_{\rm LOS}(\tau_{500})$ are necessary to reproduce the extreme area asymmetries ($\delta A \rightarrow 1$) in the observed (see Figs.~\ref{spa1t} and~\ref{spa2}) circular polarization profiles \citep{Landolfi1996}. Additionally, the inferred physical parameters could be reliant on the initial atmosphere. To avoid this dependence, we have decided to invert each individual pixel with 25 different initial atmospheric models. These models were determined by randomly perturbing the temperature stratification of the Harvard-Smithonian Reference Atmosphere (HSRA) model \citep{Gingerich1971}. The rest of the physical parameters of the initial model ($B$, $\gamma$ and ${\rm v}_{\rm LOS}$) were determined via a uniform probability distribution and considered to be independent of the optical-depth. Out of the 25 solutions obtained for a given pixel, we retain the one that yields the smallest value of $\chi^2$.

\begin{figure*}
\centering
\includegraphics[width=16cm]{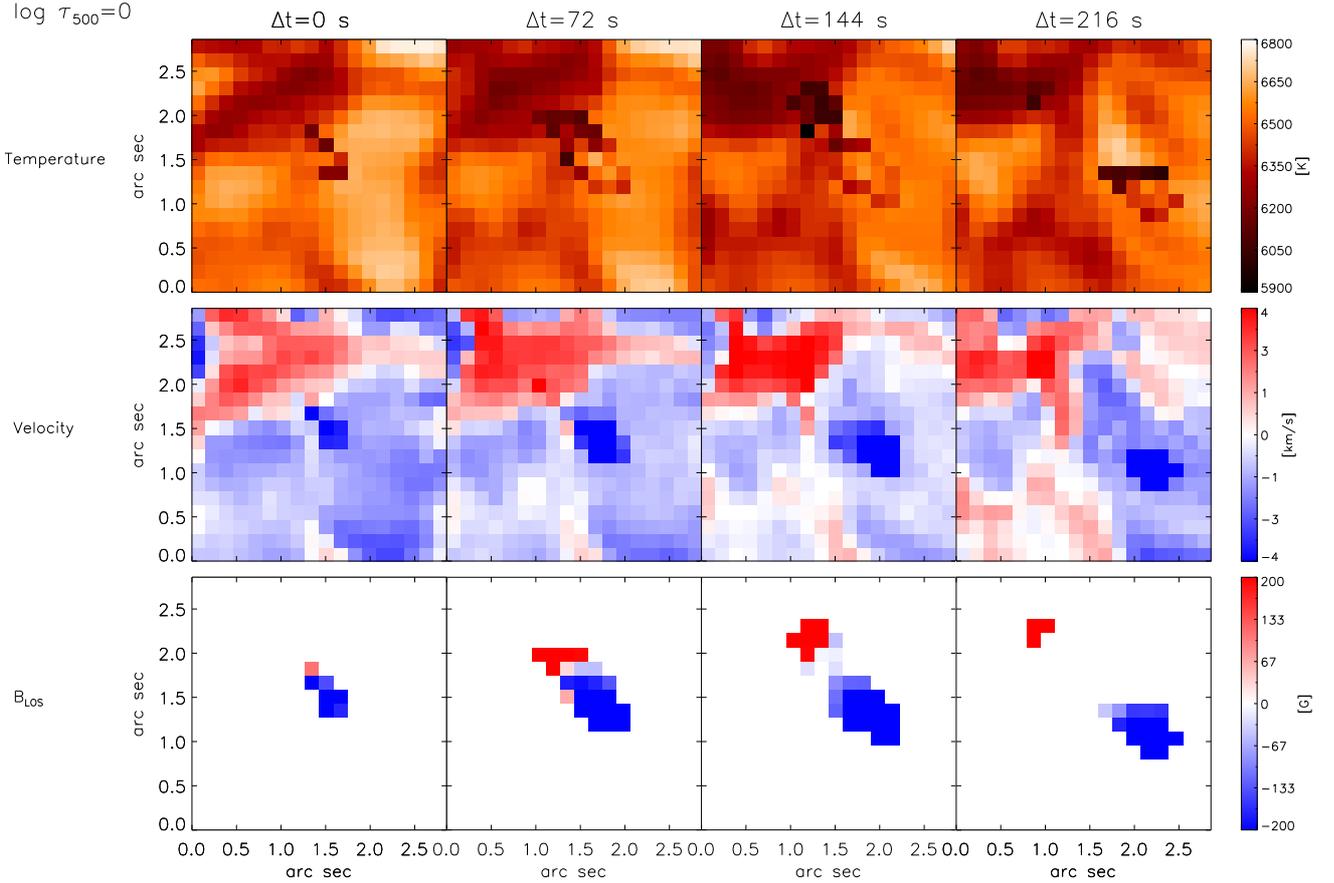}
\caption{Results from the inversion of the observed Stokes profiles in the same region as in Figure~\ref{chromos}. Each panel displays the spatial distribution of the physical parameters at an optical depth $\log\tau_{500}=0$ (i.e. continuum). {\it From top to bottom}: temperature $T$, LOS velocity ${\rm v}_{\rm LOS}$, and LOS magnetic field $B_{\rm LOS}=B \cos\gamma$. {\it From left to right}: different time steps during the evolution of the event, $\Delta t=$ 0, 72, 144, and 216 seconds.}
\label{77}
\end{figure*}

\subsection{Statistical properties}\label{stats}

We have inverted all 556 selected pixels in Hinode/SP \textit{normal} map (398 blue- and 158 red-shifted; Sect.~\ref{profiles_sec}) employing the configuration described in Sect.~\ref{conf}. Figure~\ref{mean} presents the averaged physical parameters as a function of the optical-depth evaluated at a wavelength of 500 nm ($\log\tau_{500}$) for the blue (upper row) and red cases (bottom row). The shaded-grey areas in this figure indicate the standard deviation obtained from the ensemble of pixels belonging to each of the two cases.

The mean temperature stratification (left most panels) is very similar in both blue and red cases. In the deep photosphere, $\log\tau_{500} \ge 0$, both kinds of events follow the stratification from the HSRA model (dashed lines) very closely. Above this region, $\log\tau_{500} \le 0$, the inferred temperature in both cases is larger than that of the HSRA model. However, there are two particular layers ($\log\tau_{500} \approx -3.5$ and $\log\tau_{500} \approx -2$) where the temperature in these events is the same as in the aforementioned model. The only minor difference between the inferred temperature stratifications appears close to the continuum, $\log\tau_{500} = 0$, where we obtain that the blue cases are slightly hotter (on average) than HSRA, whereas the red cases are slightly cooler. This was foreseeable because the former tend to appear at bright regions as seen on the continuum intensity, while the latter occur at dark regions (see Sect.~\ref{profiles_sec} and upper row in Fig.~\ref{chromos}).

\begin{figure*}
\centering
\includegraphics[width=16cm]{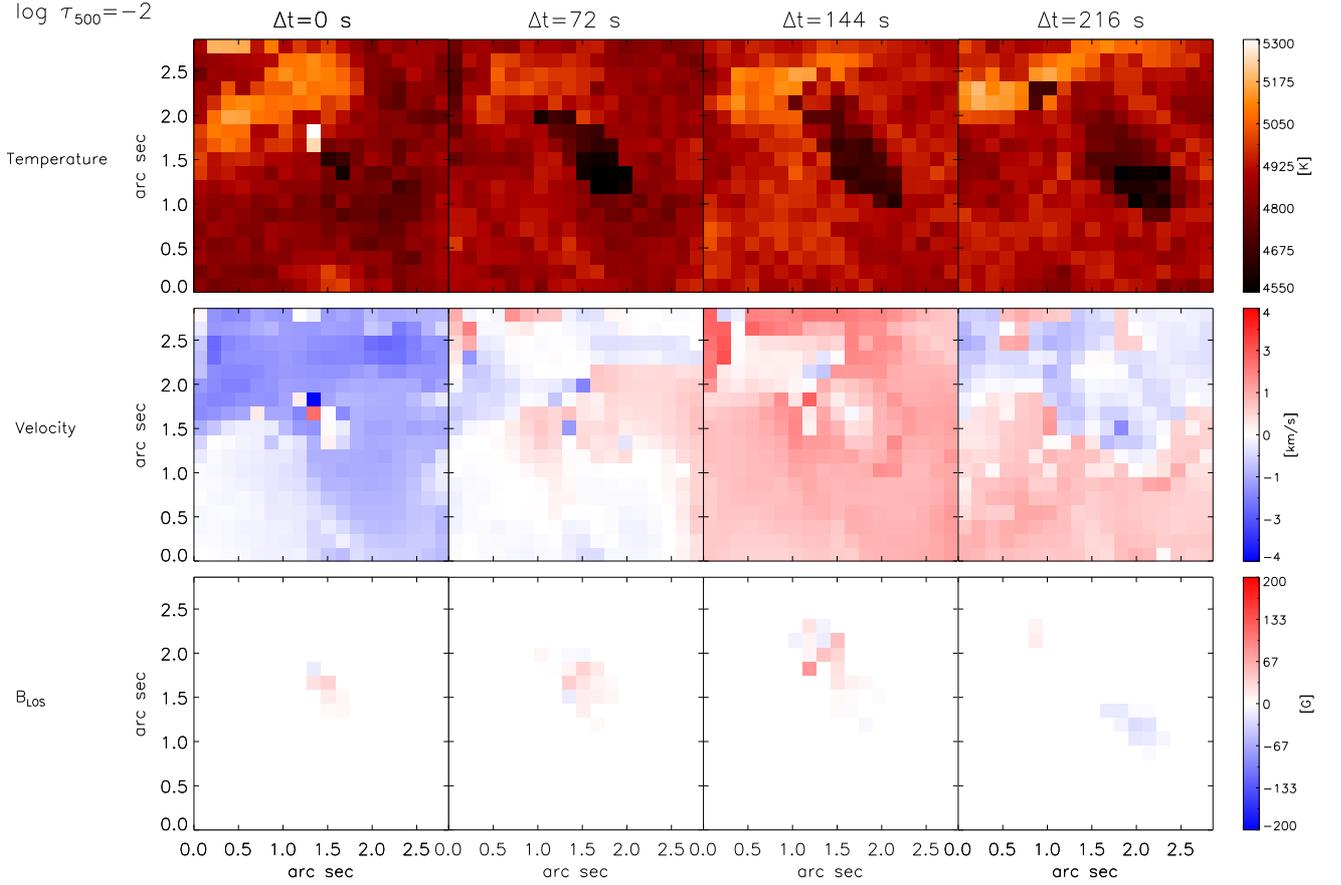}
\caption{Same as Figure \ref{77}, except showing the physical parameters at $\log\tau_{500}=-2$.} 
\label{83}
\end{figure*}

Unlike the temperature, the inferred LOS velocities (second column in Fig.~\ref{mean}) are very different in the blue and red cases. In the former, the LOS velocity changes from large blue-shifts in the deep photosphere, ${\rm v}_{\rm LOS} = -5$ km~s$^{-1}$ at $\log\tau_{500} = 0$, to red-shifts in the higher photospheric layers:  ${\rm v}_{\rm LOS} = 3$ km~s$^{-1}$ at $\log\tau_{500} = -3$. In the red cases, however, the velocity stratification is exactly opposite, with large red-shifts in the deep photosphere, ${\rm v}_{\rm LOS} = 5$ km~s$^{-1}$ at $\log\tau_{500} = 0$, which become blue-shifts at higher photospheric layers: ${\rm v}_{\rm LOS} = -3$ km~s$^{-1}$ at $\log\tau_{500} = -3$. The LOS velocities inferred from the inversion close to $\log\tau_{500} = 0$ correspond to what should have been expected from the visual inspection of the profiles (see Fig.~\ref{spa1t}), since the blue cases featured highly blue-shifted Stokes $V$ profiles and enhanced absorption on the blue wing of Stokes $I$, and vice versa for the red cases (see Sect.~\ref{profiles_sec}). 

As far as the magnetic properties of these events are concerned, the inversion returns a magnetic field strength $B$ that, despite being rather large (about 1 kG) in the deep photosphere, quickly decreases toward higher photospheric layers and becomes negligible above $\log\tau_{500} = -2$. This behavior is observed in both blue and red cases (see third column in Fig.~\ref{mean}). Meanwhile, the inclination of the magnetic field indicates that the field lines are nearly vertical close to $\log\tau_{500} = 0$, and they gradually become horizontal as we move upward in the photosphere. The value of the inclination, $\gamma$, at $\log\tau_{500} = 0$ follows the polarity of the magnetic field already seen in Fig.~\ref{chromos} (third row) and  is the opposite for blue and red cases. The inferred inclination for $\log\tau_{500} \le -2$ carries no significance because of the large scatter (see shaded areas) and because the magnetic field strength vanishes above this layer ($B \rightarrow 0$ when $\log\tau_{500} \le -2$).

\subsection{Evolution}\label{evolution}

Besides the data from the Hinode/SP {\it normal map} (Sect.~\ref{stats}), we have also inverted the Stokes profiles from the time series described in Sect.~\ref{time_evol}. This allows us to follow the evolution of the physical parameters in the example presented in Fig.~\ref{chromos}. Inferring the magnetic field vector $B$, $\gamma$, and $\phi$ from noisy data is a difficult task \citep{borrero2011} and therefore, we have decided to invert only for the temperature $T(\tau_{500})$, and LOS velocity ${\rm v}_{\rm LOS}(\tau_{500})$ in those pixels in Fig.~\ref{chromos} where the total polarization is below 2 \% of $I_{c}$. In this case, we use three nodes for the temperature and the LOS velocity. Pixels with polarization values above this threshold were treated in the way described in Sect.~\ref{stats}. The inversion of all pixels in Figure~\ref{chromos} yields the physical parameters ($T$, $B$, $\gamma$, $\phi$, and ${\rm v}_{\rm LOS}$) as a function of $(x,y,\log\tau_{500},t)$. To visualize the results, we present in Figures~\ref{77} and~\ref{83} maps of the physical parameters at two different optical depths in the photosphere: $\log\tau_{500}=0,-2$, respectively. Each of these two figures display, from the top to bottom, the temperature $T(x,y)$, LOS velocity ${\rm v}_{\rm LOS}(x,y)$, and finally the LOS magnetic field $B_{\rm LOS}(x,y)$ obtained as the product between $B$ and $\cos\gamma$. Each column presents a different time in the event's evolution: $\Delta t=0,72,144,216$ s.

We note that the temperature $T(x,y)$ at $\log\tau_{500}=0$ (upper panels in Fig.~\ref{77}) closely follows the intensity pattern (i.e., granulation) outside the strong magnetic field regions (see first and second rows in Fig.~\ref{chromos}). The Wilson depression inside the magnetic element causes the $\tau_{500}=1$ level to shift with respect to its surroundings. This effect, in  combination with the existence of vertical gradients in the temperature, produces the apparent discontinuities in the temperature between the magnetic and non magnetic region.  The color palette for ${\rm v}_{\rm LOS}(x,y)$ in the second row in Figures~\ref{77} and~\ref{83} has been chosen such that positive velocities are in red (red-shifts or downflows) and negative velocities are in blue (blue-shifts or upflows). In addition, $B_{\rm LOS}(x,y)$ (third row in Figs.~\ref{77} and~\ref{83}) shows positive polarity magnetic fields (pointing toward the observer) in red, while negative polarity magnetic fields (pointing away from the observer) are indicated in blue. 

Figure~\ref{77} demonstrates that, in the two regions with strong magnetic fields, the polarities and LOS velocities are inverted. As mentioned in Sects.~\ref{profiles_sec} and~\ref{time_evol}, these two regions correspond to the footpoints of the magnetic loop that separate from each other as time passes. Higher up in the photosphere, $\log\tau_{500}=-2$ (see Figure~\ref{83}), we can see that the temperature in the regions with strong magnetic fields are close to those of the HSRA model, where T$_{\rm HSRA}$(log $\tau_{500}$=-2)=4660 K. This is in agreement with the previous section (see also Fig.~\ref{mean}). We also observe that the LOS magnetic field and velocity above the location of the footpoints have vanished (see also the third column in Fig.~\ref{mean}). The fact that ${\rm v}_{\rm LOS}$ and $B_{\rm LOS}$ are only present at the bottom of the photosphere lead us to identify this structure as an $\Omega$-loop structure rising through the atmosphere. A more detailed analysis of the loop structure will be performed in the next section.

\section{Side view of the loop}
\label{sideview}

\subsection{Method}
\label{sideview_method}

\begin{figure}[t]
\centering
\includegraphics[width=8.0cm]{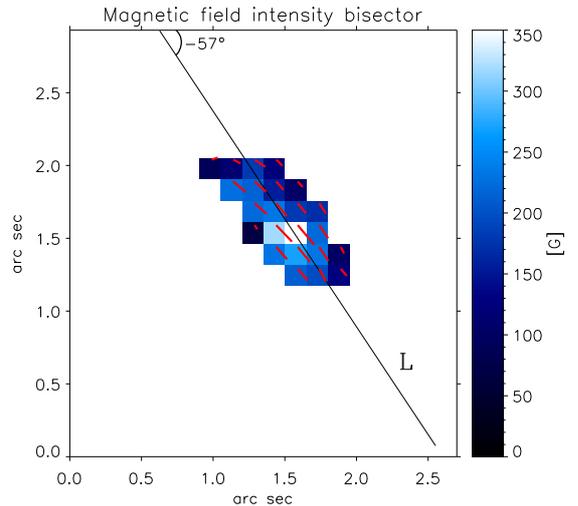}
\caption{Magnetic field intensity bisector reference, black line, for the magnetic loop analyzed using the time series. The direction along the bisector is $L$. It forms an angle of $-57^{\circ}$ with the horizontal axis. The selected time instant, $\Delta t=72$ $s$, corresponds to the second column in Figures \ref{chromos}, \ref{77} and \ref{83}. Red headless arrows display the behavior of the horizontal component of the magnetic field after solving the 180$^{\circ}$-ambiguity.}
\label{bisec}
\end{figure}

In this section, we will analyze the vertical structure of the loop in Figures~\ref{77} and~\ref{83} as seen from the side. To that end we employ the results from the inversion presented in Section~\ref{evolution}. Here we focus only on the time step $\Delta t=72$ seconds. At this time the loop is completely developed. In addition, the footpoints and the linear polarization signal that connects them are clearly discernible. To determine the physical properties of the loop as seen from the side we average, at each optical depth, the physical quantities temperature, LOS velocity, and magnetic field vector along the direction perpendicular to the line that connects both footpoints. This connecting line, or {\it bisector}, is denoted by the inclined solid black line in Figure~\ref{bisec}. In this fashion, we can transform all relevant physical parameters from the $(x,y,\tau_{500})$ coordinate system into $(L,\tau_{500})$, where $L$ refers to the direction along the bisector in Figure~\ref{bisec}.

Once in this new reference frame $(L,\tau_{500})$, we need to translate the physical parameters into a frame where the vertical axis corresponds to the geometrical height $z$ instead of the optical depth scale $\tau_{500}$. To achieve this, we employ a simplified version of the method described in \citet{klaus2010proc,klaus2010apj}. The idea is to determine a $z(\tau_{500})$ conversion for each pixel along the $L$ direction. This can be done by integrating the relation $dz=-d\tau_{500}/[\rho(\tau_{500}) \kappa(\tau_{500})]$. Here, $\kappa$ and $\rho$ correspond to the opacity and density, respectively, of the averaged ($L$-direction) atmosphere, with the latter being obtained through the assumption of hydrostatic equilibrium along the vertical $z$-axis. The solution to this equation requires an integration constant. In our case, this constant is taken as the Wilson depression $z_w=z(\tau_{500}=1)$. With this procedure, a different $z(\tau_{500})$ conversion can be established for every point $L$ along the loop. In order to establish a common $z$-scale for all pixels, we allow the aforementioned integration constant to vary along the bisector, $z_w(L)$, such that the divergence of the magnetic field, $\nabla \cdot {\rm \bf B}$, is minimized along the loop.

Before this minimization is performed, we need to solve the 180$^{\circ}$ ambiguity in the azimuth ($\phi$) of the magnetic field. This ambiguity makes it impossible to distinguish between the following two solutions: ${\rm \bf B}=(B_x,B_y,B_z)$ and ${\rm \bf B}^{\dagger}=(-B_x,-B_y,B_z)$. If left unattended, the terms corresponding to the horizontal derivatives in $\nabla \cdot {\rm \bf B}$ ($\partial B_x/\partial x$ and $\partial B_y/\partial y$) will dominate over the vertical derivative ($\partial B_z/\partial z$), thereby hindering the minimization of the divergence of the magnetic field by changing the boundary condition $z_w(L)$. The ambiguity is solved by choosing, at each pixel along the bisector in Fig.~\ref{bisec}, either ${\rm \bf B}$ or ${\rm \bf B}^{\dagger}$, such that the magnetic field in the $XY$-plane always points along the same quadrant. The red headless arrows in Fig.~\ref{bisec} show the direction of the magnetic field on the horizontal plane after the correction for the 180$^{\circ}$-ambiguity, illustrating that the field lines connect both footpoints of the structure.

\begin{figure*}
\centering
\includegraphics[width=15.0cm]{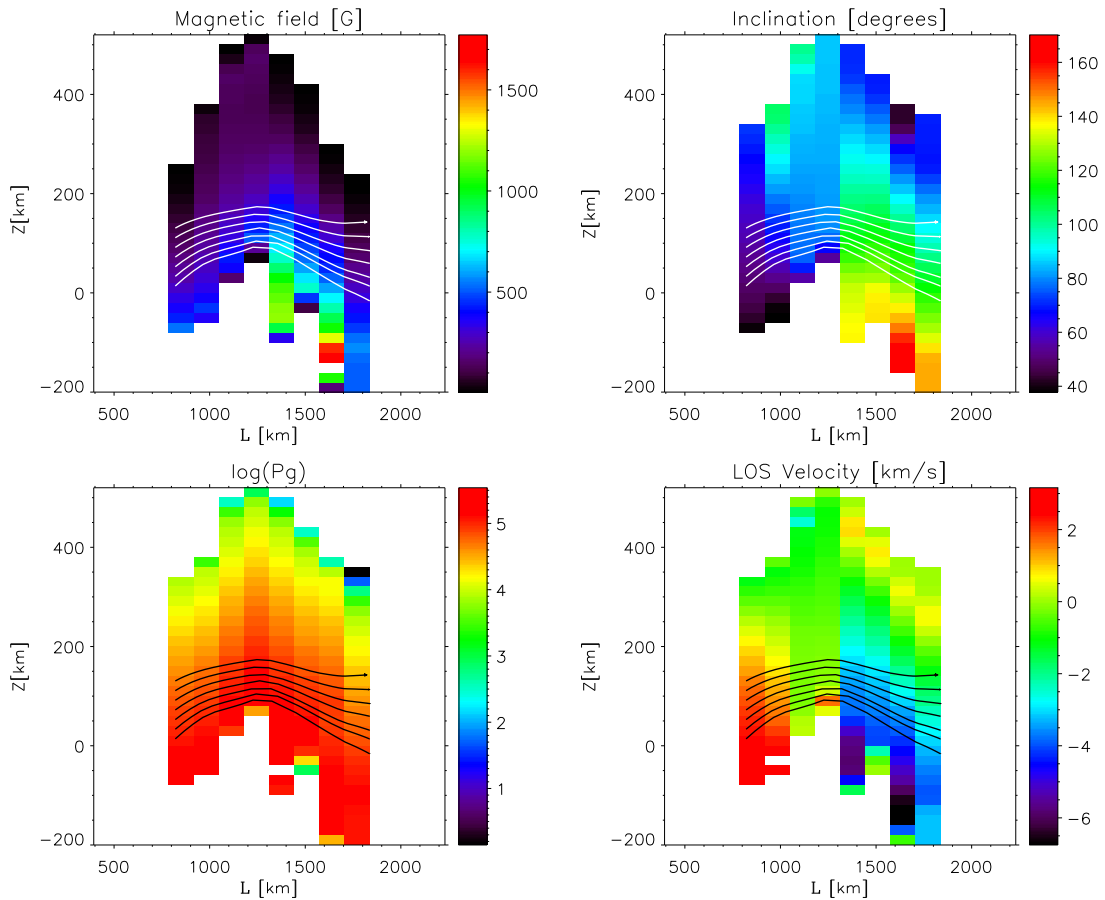}
\caption{Side view of the observed magnetic loop. {\it Top left panel}: magnetic field $B(L,z)$. {\it Top right panel}: inclination of the magnetic field $\gamma(L,z)$. {\it Bottom left panel}: logarithm of the gas pressure $P_g(L,z)$. {\it Bottom right panel}: LOS component of the velocity ${\rm v}_{\rm LOS}(L,z)$. White/black lines trace the magnetic field lines.}
\label{loop80}
\end{figure*}

\subsection{Results}
\label{sideview_results}

After a common $z$ scale has been established for all pixels along the bisector in Fig.~\ref{bisec}, we can study the behavior of the physical parameters as a function of the common vertical $z$-coordinate and the distance $L$ along the loop. This is equivalent to looking at the structure from the side, hence the title of Section~\ref{sideview}. Figure \ref{loop80} displays: magnetic field $B(L,z)$ (upper left panel), inclination of the magnetic field $\gamma(L,z)$ (upper right panel), gas pressure $P_g(L,z)$ (lower left panel; logarithmic scale), and LOS velocity ${\rm v}_{\rm LOS}(L,z)$ (lower right panel). Each panel indicates the magnetic field lines of the loop displayed in white or black. We can see that the minimization of $\nabla \cdot {\rm \bf B}$ (which shifts up or down the vertical scale at each pixel along the bisector; see Sect.~\ref{sideview_method})  forces the central sections of the loop to be located higher in the photosphere than the footpoints. This behavior produces a clear $\Omega$-loop shape: the footpoints (formed deep in the photosphere) are characterized by somewhat vertical fields with opposite polarities: $\gamma \approx 45^{\circ}$ on the left footpoint ($L \approx 700$ km), but $\gamma \approx 120^{\circ}$ on the right footpoint ($L \approx 1600$ km). Meanwhile, the apex of the loop ($L \approx 1200$ km) possesses a more horizontal magnetic field ($\gamma \approx 90^{\circ}$). In addition, the LOS velocity ${\rm v}_{\rm LOS}$ at each footpoint has the opposite sign (bottom-right panel), thus indicating that the plasma flows from the right footpoint (upflow; ${\rm v}_{\rm LOS} < 0$) toward the left footpoint (downflow; ${\rm v}_{\rm LOS} > 0$). We emphasize that the values of ${\rm v}_{\rm LOS}$ at the footpoints are of the order of $3-4$ km~s$^{-1}$, but if we project these velocities along the loop's axis, assuming that the velocity and magnetic field vectors are parallel, we obtain absolute values of $5-6$ km~s$^{-1}$ at the footpoints. These velocities are significantly faster than the typical convective velocities seen in the granulation.

The bottom left panel in Fig.~\ref{loop80} allows us to detect gas pressure imbalances $\Delta P_g$ between two points along the loop at a fixed height. This panel should be treated with caution however, because of the exponential dependence of $P_g$ with height. This dependence means that small errors in the determination of the Wilson depression at each point along the loop, $z_w(L)$, greatly increase the uncertainty in the gas pressure difference between any two given points.

\begin{figure}
\centering
\includegraphics[width=8.0cm]{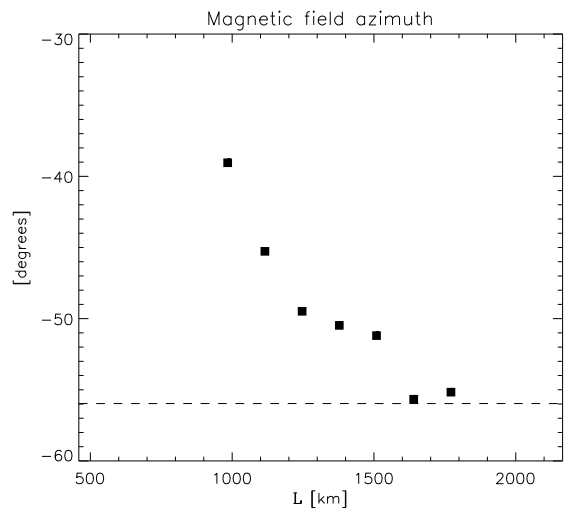}
\caption{Magnetic field azimuth. Squares mark the mean value of the inverted pixels and the dotted line designates the orientation of the black line plotted in Figure \ref{bisec}, i.e., the bisector.} 
\label{azimuth}
\end{figure}

During the inversion, only one node was given to the azimuthal angle of the magnetic field $\phi$ (see Sect.~\ref{conf}). Because of this, $\phi$ does not depend on $z$ (or $\tau$). However, the azimuth angle does feature an interesting behavior along the loop. This behavior, $\phi(L)$, is indicated in Figure~\ref{azimuth}, where we observe that the average magnetic field is not completely aligned with the structure. In the upflowing footpoint (${\rm v}_{\rm LOS} < 0$; $L \approx 1600$ km), the azimuth of the field closely follows the inclination of the bisector ($-57^{\circ}$ in Fig.~\ref{bisec}), but it gradually deviates from this value as we approach the downflowing footpoint (${\rm v}_{\rm LOS} > 0$; $L \approx 700$ km). We conclude, therefore, that the magnetic field shows a certain degree of twist along the loop. This can be produced by a structure with magnetic helicity. The torsion of the magnetic field lines is also visible in the relative orientation between the bisector and the red headless arrows in Figure \ref{bisec}.

\section{Discussion}

The analysis of the data from the Hinode/SP time-series revealed two footpoints with opposite polarities connected by a region that harbors clear linear polarization signals ($> 0.01~I_{c}$; see Sects~\ref{profiles_sec}, ~\ref{time_evol}, and Fig.~\ref{chromos}). Because of this, it is tempting to ascribe the inferred configuration of the magnetic field to a magnetic loop. The analysis of the horizontal component of the magnetic field shows that the field lines go from one footpoint to the other (Fig.~\ref{bisec}), albeit with some degree of twist (Fig.~\ref{azimuth}). Furthermore, by imposing the  $\nabla \cdot {\rm \bf B} = 0$ condition, we have been able to determine (see Sect.~\ref{sideview}) that the linear polarization signals (corresponding to horizontal magnetic fields) come from higher photospheric layers than those responsible for the circular polarization signals (associated with more vertical magnetic fields) of opposite polarity at the footpoints. All these results lend support to a configuration of the magnetic field in the form of an $\Omega$-loop. This kind of topology has already been reported by \cite{MartinezGonzalez2007,Centeno2007,MartinezGonzalez2009,Ishikawa2010}. A similar type of configuration, including the presence of single-lobed Stokes $V$ profiles, has also been described in \cite{Viticchie2012}.

Once we have established the configuration of the magnetic field, we turn our attention to the inferred velocities. The fact that the LOS velocities are also opposite at the loop footpoints (see the second column in Fig.~\ref{mean} and the bottom right panel in Fig.~\ref{loop80}) indicates that the magnetized plasma flows from one footpoint towards the other. In spite of not finding any clear gas pressure imbalance between the footpoints (see Sect.~\ref{sideview_results}), we believe that the siphon flow mechanism \citep[][ and references therein]{Montesinos1993} is the only physical mechanism (of those listed in Sect.~1) that explains the observed flow along this magnetic $\Omega$-loop (i.e., the observed kinematic and magnetic properties).

There are, however, some features of the inferred physical parameters that do not entirely match the above scenario. We refer in particular to the strong increase in the temperature in the mid-photosphere at the footpoints of the loop (see leftmost panels in Fig.~\ref{mean}), and the change in the sign (also in the mid-photosphere) of the LOS component of the velocity ${\rm v}_{\rm LOS}$ where the magnetic field drops to zero (second and third columns in Fig.~\ref{mean}). It is possible to argue, however, that these two features appear as a consequence of the limitations in our modeling of the solar atmosphere. In the following, we will elaborate on this particular point.

\begin{figure*}
\centering
\includegraphics[width=17cm]{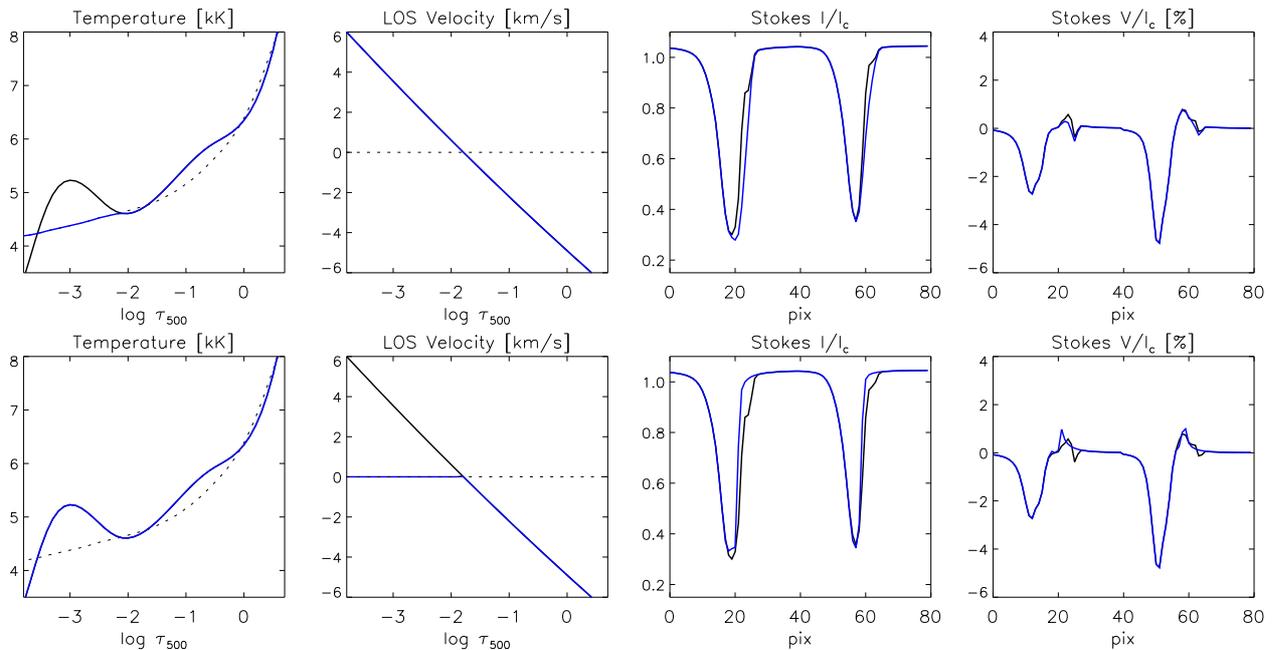}
\caption{Synthetic Stokes profiles using a temperature stratification without the inferred increased at high layers, first row. The same study but changing to zero the LOS velocity at high layers is presented in the second row. Black designates the original profiles obtained using the mean atmosphere showed in blue in Figure \ref{mean} while blue lines mark the newer results.}
\label{sint}
\end{figure*}

In Section~\ref{conf} we mentioned that, according to \cite{Landolfi1996}, several nodes in the magnetic field and LOS velocity were needed to reproduce the large area asymmetries in the observed Stokes $V$ profiles (see Fig.~\ref{spa1t}). As a result, we have inferred an optical depth dependence of the LOS velocity where ${\rm v}_{\rm LOS}(\tau_{500})$ changes from large blue-shifts in the deep photosphere ($\log\tau_{500} \approx 0$) to large red-shifts in the higher layers ($\log\tau_{500} \approx -3$) at the upflowing footpoint, and vice versa at the downflowing footpoint. The question is therefore whether this change in the sign of the velocity in the high photosphere (and also the temperature enhancement) is needed to reproduce the asymmetries in the observed circular polarization profiles. To investigate this we have performed two different experiments.

In the first experiment,, we took the optical depth dependence of the physical parameters inferred from the inversion of the blue cases (upper row in Fig.~\ref{mean}) and proceeded to calculate the emerging Stokes profiles from this mean atmospheric model using the synthesis module of SIR. The results are represented by solid black lines in the upper panels of Figure~\ref{sint}. As expected, the circular polarization profiles are highly asymmetric ($\delta A \rightarrow 1$) and very similar to the blue Stokes profiles in Figs.~\ref{spa1t} and~\ref{spa2} (as they should because the atmospheric model was actually obtained from the inversion of those profiles). We then perform a synthesis using those very same physical parameters, but substituting the temperature enhancement above $\log\tau_{500} < -2$ for the temperature stratification of the HSRA model \citep{Gingerich1971}. The new atmosphere and resulting Stokes profiles are indicated in blue in the upper row of Fig.~\ref{sint}. As can be seen, Stokes $V$ has changed very little after modifying the model's temperature stratification. Stokes $I$, however, does show some differences.

The second experiment is identical to the first, except now we maintain the temperature stratification $T(\tau_{500})$ of the original model (see leftmost bottom row in Fig.~\ref{sint} and upper left panel in Fig.~\ref{mean}). This time, however, we neglect the change of sign in the LOS velocity in the upper layers, and, instead, we set ${\rm v}_{\rm LOS}=0$ above $\log\tau_{500} < -2$ (see second column bottom panel in Fig.~\ref{sint}). Then we perform a synthesis to calculate the emergent Stokes profiles from this modified atmosphere. The results are indicated in blue  in the bottom row of Fig.~\ref{sint}. Again, while Stokes $I$ shows some differences, Stokes $V$ is almost the same as in the original model.

The same two experiments except using, as an original atmosphere, the red case (bottom row in Fig.~\ref{mean}) instead of the blue case (upper row in Fig.~\ref{mean}), lead to identical results. From these tests we can conclude that the neither the temperature enhancement nor the velocity change in the higher photospheric layers are needed to reproduce the asymmetry of the observer circular polarization profiles. Interestingly, they are needed to reproduce Stokes $I$ instead. Once the effect of these gradients above $\log\tau_{500} < -2$ on the asymmetries of Stokes $V$ have been ruled out, there is no particular reason as to why the atmosphere in this particular photospheric layer must be really there. In fact, at this point, this scenario is indistinguishable from one in which we extract the physical parameters above $\log\tau_{500} < -2$ in Fig.~\ref{mean}, and place them in a spatially unresolved second component located at the same height as the footpoint. This second component would be unmagnetized and carry a downflow associated with the upflowing footpoint (blue pixels in Fig.~\ref{spa1t}). However, next to the downflowing footpoint (red pixels in Fig.~\ref{spa1t}), this second component  would carry an unmagnetized upflow. Its role would be limited to fitting, as we just described above, Stokes $I$, in particular its wings (see Sect.~\ref{profiles_sec}).

Independent of which of the two aforementioned scenarios is correct (non magnetic atmosphere above the loop's footpoints, or at the same height), it does not seem coincidental that this additional component harbors opposite LOS velocities than those of the footpoints. Unfortunately, with the current observations it is not possible to distinguish between these two possibilities, nor to study the nature of this second component and whether it interacts with the magnetic $\Omega$-loop.

\section{Conclusions}

The inspection of circular polarization signals at $\Delta\lambda=\pm272$~m{\AA} reveals the presence of pixels with highly blue-shifted single-lobed Stokes $V$ profiles grouped in patches of about 6.2$\pm$3 pixels. Very often (in 73 \% of the detected cases), these patches are accompanied by highly red-shifted single-lobed Stokes $V$ profiles that occupy a smaller, 3.5$\pm$2.1 pixels, area. In the remaining cases, 27 \% of the total, the patches of highly blue-shifted single-lobed Stokes $V$ profiles are isolated from other magnetic structures. We have referred to these patches as \emph{blue} and \emph{red} cases. The polarization profiles observed in these events are such that Stokes $I$ features a strongly blended blue or red wing in the blue or red cases, respectively. Although the amplitude of the circular polarization profiles is always higher in the blue cases than in the red cases, the sign of their only existing lobe is the same. These two kinds of events appear within 2\arcsec of each other. Interestingly, whenever they are very close, the pixels in between them display enhanced linear polarization signals. In this region of enhanced linear polarization, Stokes $V$ has a shape that closely resembles that of Stokes $Q$. 

The study of the continuum intensity map $I_{c}$ shows that blue cases appear over bright regions, usually located at the granular edges, while red cases are placed in dark intergranular lanes. Overlaying the events to a magnetogram reveals that most of them occur around magnetic field concentrations (i.e., network), and almost never in the quiet-Sun internetwork.

The inversion of the Stokes profiles, detected as blue or red cases in the \emph{normal} map, allowed us to retrieve the optical depth stratification of the physical parameters through the solar photosphere: $T(\tau_{500})$, $B(\tau_{500})$, $\gamma(\tau_{500})$, ${\rm v}_{\rm LOS}(\tau_{500})$, etc. We have found that the temperature at $\log\tau_{500} = -1$ and $-3$ is larger than in the HSRA model (average quiet-Sun model), but lower than that model at $\log\tau_{500} = -2$. This happens in both blue and red cases, and the only difference between them at $\log\tau_{500} =0$, is that blue/red cases are slightly hotter/cooler than HSRA model. This last feature is just a consequence of the blue cases having larger continuum intensities than the red cases. The magnetic field strength $B$ in both types of events also behaves very similarly, with values of up to 1 kG in the deep photosphere ($\log\tau_{500} = 0$), which quickly decrease upward until vanishing at $\log\tau_{500}=-2$. The inclination of the magnetic field $\gamma$, on the other hand, indicates that the magnetic field has opposite polarities in these events, that is to say, if $\gamma < 90^{\circ}$ for blue case, then $\gamma > 90^{\circ}$ for the associated red case (or vice versa). The velocities also display a rather different behavior. In the blue cases, the LOS velocity ${\rm v}_{\rm LOS}$ changes from large blue-shifts in the deep photosphere ($\log\tau_{500} = 0$) to large red-shifts in the upper photosphere ($\log\tau_{500} = -3$). In the red cases, we observe exactly the opposite, that is, large red-shifts in the deep photosphere that turn into large blue-shifts higher up.

The inversions of the Stokes profiles, detected as blue or red cases in the time series, revealed the same properties of the physical parameters as in the \emph{normal} map. With the time series, however, we have been able to analyze the evolution of these events and discovered that their magnetic field configuration is that of an $\Omega$-loop, where the footpoints move further from each other as time passes. The physical process that best explains the observed configuration of the magnetic field while, at the same time, accounting for the high velocities at the footpoints, is a siphon flow. However, we were not able to detect the required gas pressure or magnetic field imbalance between footpoints. Moreover, although we have argued that it might be an artefact from the geometrical model we have defined prior to our inversions, the large velocity gradients at each of the footpoints, are difficult to reconcile with the picture of a siphon flow along a magnetic arch. Thus, at this point we are not able to fully establish the origin or physical mechanism responsible for these events.

Data with better spatial resolution, or including spectral lines that convey more reliable information on the higher layers of the solar photosphere, could help in the interpretation of the phenomena described in this paper. Hopefully, this will be possible with future ground-based facilities such as EST \citep{Collados_EST} and ATST \citep{Keil_ATST}, or from satellites like Solar-C \citep{Suematsu_SolarC}. It would be also interesting to search for these events into realistic MHD simulations to find out whether there is some kind of relation between their emergence and what occurs in the surrounding medium \cite[see][]{danilovic2014}.

￼\begin{acknowledgement}
C.Q.N. thanks to the researchers of the Kiepenheuer-Insitut f$\ddot{\rm u}$r Sonnenphysik (Freiburg, Germany) for their insightful comments that helped to develop this work. C.Q.N. also thanks to the director Oskar von der L$\ddot{\rm u}$he for the opportunity to have a wonderful stay in the KIS Institute, the same stay that seeds this paper. This work has been funded by the Spanish MINECO through Projects No. AYA2009-14105-C06-03 and AYA2011-29833-C06. The data used here were acquired in the framework of Hinode Operation Plan 14 (Hinode-Canary Islands joint campaign). \textit{Hinode} is a Japanese mission developed and launched by ISAS/JAXA, with NAOJ as a domestic partner, and NASA and STFC (UK) as international partners. It is operated by these agencies in cooperation with ESA and NSC (Norway). Extensive use of the NASA Astrophysical Data System has been made.
￼\end{acknowledgement}

\bibliographystyle{aa} 
\bibliography{blue.bib} 

\end{document}